\documentclass[11pt,a4paper]{article}
\usepackage{fullpage}
\usepackage{amsmath, amsthm, amsfonts,  amssymb, latexsym}
\usepackage[colorlinks, citecolor=Red, linkcolor=Blue, urlcolor=blue]{hyperref}
\usepackage{caption}
\usepackage[dvipsnames]{xcolor}
\usepackage{tikz}
\usepackage{pgfplots}
\usetikzlibrary{arrows,arrows.meta, decorations.pathmorphing, decorations.markings, decorations.pathreplacing,intersections, pgfplots.fillbetween,patterns,shadings}
\usepackage{mcite}
\usepackage{enumerate}
\usepackage{cancel}

\newcommand{\lhs}{l.h.s.\ }
\newcommand{\wrt}{w.r.t.\ }
\newcommand{\cf}{cf.\ }

\makeatletter
\DeclareRobustCommand{\ibar}{\mathord{%
  \text{$\m@th\mkern-2mu\raisebox{-0.7ex}[0pt][0pt]{$\mathchar'26$}\mkern-7mu i$}%
}}
\makeatother

\newcommand{\ud}{\mathrm{d}}
\newcommand{\del}{\partial}

\newcommand{\eps}{\varepsilon}

\newcommand{\rI}{{\mathrm{I}}}
\newcommand{\rII}{{\mathrm{II}}}
\newcommand{\rIII}{{\mathrm{III}}}
\newcommand{\rIV}{{\mathrm{IV}}}

\newcommand{\const}{{\mathrm{const}}}

\newcommand{\beq}{\begin{equation}}
\newcommand{\eeq}{\end{equation}}

\DeclareMathOperator{\sign}{sign}

\begin{document}

\title{Dynamics of spherical charged dust shells in de Sitter space}
\author{Christoph Giese\thanks{cg96dadu@studserv.uni-leipzig.de} \ and Jochen Zahn\thanks{jochen.zahn@itp.uni-leipzig.de} \\ Institut f\"ur Theoretische Physik, Universit\"at Leipzig\\ Br\"uderstr.\ 16, 04103 Leipzig, Germany}

\date{September 5, 2022}

\maketitle

\begin{abstract}
We study the dynamics of charged, spherically symmetric dust shells in the presence of a positive cosmological constant. We find generalizations of the well-known solutions in asymptotically flat spacetime, in particular orbits into ``parallel universes'', but also new solutions corresponding to a ``bounce'' of the shell before an event horizon has formed. We also discuss ``bubble'' solutions, in which a charged shell and an oppositely charged singularity are spontaneously created and annihilated.
\end{abstract}

\section{Introduction}

The study of the dynamics of thin shells in General Relativity has a long history, dating back to the work of Lanczos \cite{Lanczos1924} and later Israel \cite{Israel1966}. The aim of such models is typically not to provide realistic solutions, but rather as yielding simple analytically tractable toy models, in particular for gravitational collapse. In this spirit, Boulware \cite{Boulware1973} studied the dynamics of spherically symmetric charged thin dust shells in asymptotically flat spacetime, showing in particular the existence of solutions for which the shell crosses the Cauchy horizon, and which, depending on the parameters, may either hit the singularity of Reissner--Nordstr\"om spacetime or evade it and escape into another asymptotically flat ``parallel universe''.

Beyond a Cauchy horizon, which exists in rotating or charged stationary black holes, the evolution of (test) fields is no longer determined by their initial data, undermining the predictivity of the theory. Penrose \cite{Penrose} argued that a Cauchy horizon should be unstable, i.e., become singular under generic fluctuations. There are several mathematically precise formulations of this \emph{strong cosmic censorship conjecture}, and while some have been shown to hold on Reissner-Nordstr\"om spacetime \cite{Dafermos2003, LukOh2015}, others do not hold \cite{Dafermos2003}, \cf also \cite{DiasReallSantos2018} for an overview. The basic mechanism, already identified by Penrose, is the blue-shift of perturbations near the Cauchy horizon. This effect can be countered by the red-shift due to cosmological expansion in the presence of a positive cosmological constant. As a consequence, the most popular formulation of strong cosmic censorship, due to Christodoulou \cite{Christodoulou},
can be violated for near-extremal Reissner--Nordstr\"om--de Sitter (RNdS) spacetimes \cite{HintzVasy, CardosoEtAl}.

While it was subsequently shown that the expectation value of the stress tensor of a quantum field in RNdS (in any state that naturally arises in a gravitational collapse scenario) diverges in a way that restores strong cosmic censorship \cite{HWZ, HKZ} (see also \cite{Ori2019} for related earlier work), it was also argued \cite{Hod2019} that spacetimes that violate the strong cosmic censorship conjecture can not arise in a dynamical collapse, due to quantum (gravity) effects. We take this as a motivation to study the dynamics of spherical charged thin dust shells in the presence of a positive cosmological constant. One goal is to see whether spacetimes admitting for a violation of strong cosmic censorship can arise in this way. More generally, we are interested in the effect of the cosmological constant on the dynamics, i.e., whether the solutions found by Boulware \cite{Boulware1973} are stable under turning on the cosmological constant, and whether there are also orbits that qualitatively differ from the ones present in the asymptotically flat setting.

Our main results are as follows: The solutions found by Boulware \cite{Boulware1973} can indeed be naturally seen as limits of solutions in the presence of a cosmological constant $\Lambda$. In particular, for any set of parameters $m$ (mass), $Q$ (charge) and $\Lambda$ of a black hole solution, there is a collapse solution leading to a RNdS spacetime with these parameters. Furthermore, there are qualitatively new solutions in which the shell ceases to collapse and expands again before crossing the event horizon. Similar solutions were already found for uncharged shells in the presence of a cosmological constant \cite{YamanakaEtAl}. As we are slightly more general in our assumptions than \cite{Boulware1973}, we also find solutions corresponding to spontaneous creation, followed by expansion, contraction, and annihilation, of a charged shell in de Sitter spacetime. Finally, we prove that, as for a vanishing cosmological constant \cite{Boulware1973}, collapse to a naked singularity is only possible for negative rest mass of the shell.

\section{Setup}

\begin{figure}
\centering
\begin{tikzpicture}[scale=0.75]
\begin{axis}[
    axis lines = left,legend style={draw=none},
    xlabel = $|Q| \kappa$,
    ylabel = {$m \kappa$},xmax=0.295,ymax=0.28,
	scaled ticks=false, tick label style={/pgf/number format/.cd, fixed},
	legend pos=south east]
\addplot[blue,domain=0:0.288675,samples=50,line width=1pt,name path=A]{(1/6*(1+(1-12*x^2)^(1/2)))^(1/2)*(1-2*(1/6*(1+(1-12*x^2)^(1/2))))};
\addlegendentry{$r_+ =r_c$};
\addplot[green,domain=0:0.288675,samples=50,line width=1pt,name path=B]{(1/6*(1-(1-12*x^2)^(1/2)))^(1/2)*(1-2*(1/6*(1-(1-12*x^2)^(1/2))))};
\addlegendentry{$r_+ =r_-$};
\node[circle,fill,inner sep=1pt] at (axis cs:0.288675,0.2721655) {};
\addplot[gray!10] fill between[of=A and B];
\end{axis}
\end{tikzpicture}
\caption{In grey the parameter space corresponding to the generic black hole case $r_- < r_+ < r_c$. The blue and green lines correspond to the extremal naked singularity and the extremal black hole case, respectively. The black circle represents the ultra-extreme case. The remaining parameter space is the generic singularity case.}
\label{fig:ParameterSpace}
\end{figure}
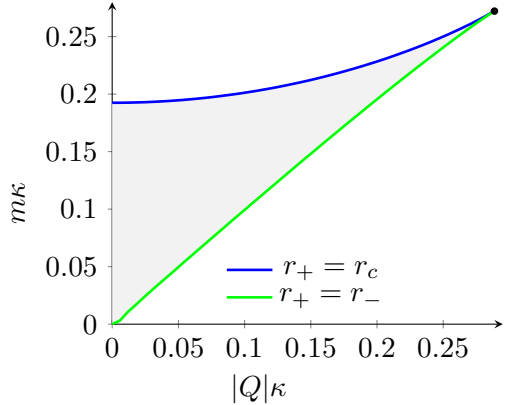

We use units such that $c = G = 1$. We consider a spherical thin shell of total charge $Q$ separating a de Sitter (dS) region from a RNdS region. The metric is in both cases of the form
\beq
\label{eq:metrics}
 g_\pm = - f_\pm(r) \ud t^2 + f_\pm(r)^{-1} \ud r^2 + r^2 \ud \Omega^2
\eeq
with $\ud \Omega^2$ the metric on the unit sphere. For the de Sitter region
\beq
 f_-(r) = 1 - \kappa^2 r^2,
\eeq
where $\kappa = \sqrt{\Lambda / 3}$ with $\Lambda$ the cosmological constant, while for the RNdS region
\beq
 f_+(r) = 1 - \frac{2m}{r} + \frac{Q^2}{r^2} - \kappa^2 r^2,
\eeq
where $m > 0$ will always be assumed. Without loss of generality, we can also assume $Q \geq 0$. While $f_-(r)$ only has a single root $r_H = \kappa^{-1}$ (corresponding to the Hubble radius) on the positive axis, $f_+(r)$ may have up to three positive roots (from the form of $f_+(r)$ it follows that the sum of the roots vanishes, so there can be at most three positive ones). One then distinguishes the following cases \cite{BrillHayward}: In the \emph{generic black hole case} $f_+(r)$ has three positive roots $r_- < r_+ < r_c$, called the Cauchy (or inner) horizon, the event horizon, and the cosmological horizon. The cases when two or all of the roots coincide are called the \emph{extremal black hole case} ($r_- = r_+$), the \emph{extremal naked singularity case} ($r_+ = r_c$), and the \emph{ultra-extreme case} ($r_- = r_+ = r_c$). When two of the roots become complex and thus only a single positive root is left, one interprets the latter as the cosmological horizon and calls this the \emph{generic naked singularity case}. The two extremal cases can be parameterized as \cite{BrillHayward,Romans1992}
\begin{align}
 m_\pm & = P_\pm \left( 1 - 2 \kappa^2 P_\pm^2 \right), & P_\pm^2 & = \frac{1}{6 \kappa^2} \left( 1 \pm \sqrt{ 1 - 12 \kappa^2 Q^2 } \right),
\end{align}
see Fig.~\ref{fig:ParameterSpace}. 

In this work, we will focus on the generic black hole case $r_- < r_+ < r_c$. Obviously, $f_+(r)$ is negative for $r > r_c$. Hence, it must be positive for $r_+ < r < r_c$, negative for $r_- < r < r_+$ and again positive for $0 < r < r_-$. We call these regions $\rIV$, $\rI$, $\rII$, and $\rIII$, respectively.
Similarly, $f_-(r)$ is positive for $0 < r < r_H$ and negative for $r > r_H$. We call these regions $\tilde \rI$ and $\tilde \rII$, respectively.

We also note that in the generic black hole case, $r_c < r_H$, as one can see as follows: For $r_c > r_H$, we would need $f_+(r_H) > 0$ (by continuity from $Q = 0$ and small enough $m$, we can assume that $r_+ < r_H$), which implies that $2 m \kappa < \kappa^2 Q^2$. One easily checks that this curve is always below the green curve in Fig.~\ref{fig:ParameterSpace}. 

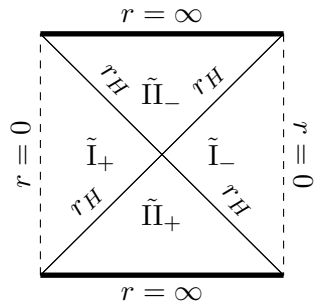
\begin{figure}
\centering
\begin{tikzpicture}[scale=0.8]
\draw (0,0)--(4,4);
\draw[dashed] (4,4)--(4,0)node[midway,above, sloped] (TextNode) {$r=0$};
\draw [line width=2pt](0,0)--(4,0)node[midway,below, sloped] (TextNode) {$r=\infty$};
\draw[dashed] (0,0)--(0,4)node[midway,above, sloped] (TextNode) {$r=0$};
\draw (0,4)--(4,0);
\draw (0,0)--(2,2)node[midway,above, sloped] (TextNode) {$r_H$};
\draw (2,2)--(4,4)node[midway,above, sloped] (TextNode) {$r_H$};
\draw (0,4)--(2,2)node[midway,above, sloped] (TextNode) {$r_H$};
\draw (2,2)--(4,0)node[midway,above, sloped] (TextNode) {$r_H$};
\draw [line width=2pt](0,4)--(4,4)node[midway,above, sloped] (TextNode) {$r=\infty$};
\node at (2,1){$\tilde \rII_+$};
\node at (2,3){$\tilde \rII_-$};
\node at (1,2){$\tilde \rI_+$};
\node at (3,2){$\tilde \rI_-$};
\end{tikzpicture}
\caption{Maximally extended de Sitter spacetime. The dashed lines indicate the origin of coordinates and the thick lines conformal infinity.}
\label{fig:dS}
\end{figure}

The above metrics have coordinate singularities at the roots of $f_\pm$. By choosing suitable adapted coordinates, the metrics can be analytically continued beyond the roots. A conformal diagram of maximally extended dS is shown in Fig.~\ref{fig:dS}. As indicated above, regions $\tilde \rI_\pm$ cover the range $r \in (0, r_H)$, while $\tilde \rII_\pm$ cover the range $r \in (r_H, \infty)$. 
Note that $\del_t$ is timelike in region $\tilde \rI_\pm$ but spacelike in $\tilde \rII_\pm$. Also note that, analogously to \cite{Boulware1973}, we distinguish the regions by a further index $\pm$. For example, in region $\tilde \rI_+$, the radius $r$ increases to the right, whereas it decreases to the right in $\tilde \rI_-$. This will be useful later in the discussion of shell trajectories.

\begin{figure}
\centering
\begin{tikzpicture}[scale=0.8]
\draw (0,0)--(2,2) node [midway, below, sloped] (TextNode) {$r_-$};
\draw (2,2)--(4,0)node [midway, below, sloped] (TextNode) {$r_-$};
\draw (2,2)--(4,4)node [midway, above, sloped] (TextNode) {$r_-$};
\draw (2,2)--(0,4)node [midway, above, sloped] (TextNode) {$r_-$};
\draw (0,4)--(2,6)node [midway, below, sloped] (TextNode) {$r_+$};
\draw (2,6)--(4,4)node [midway, below, sloped] (TextNode) {$r_+$};
\draw [line width=2pt] (0,0) -- (0,4) node [midway, above, sloped] (TextNode) {$r=0$};
\draw [line width=2pt] (4,4) -- (4,0) node [midway, above, sloped] (TextNode) {$r=0$};
\draw (4,4)--(6,6) node [midway, above, sloped] (TextNode) {$r_c$};
\draw (4,8)--(6,6) node [midway, below, sloped] (TextNode) {$r_c$};
\draw (4,8)--(2,6)node [midway, above, sloped] (TextNode) {$r_+$};
\draw (-2,6)--(0,4) node [midway, above, sloped] (TextNode) {$r_c$};
\draw (-2,6)--(0,8) node [midway, below, sloped] (TextNode) {$r_c$};
\draw (0,8)--(2,6)node [midway, above, sloped] (TextNode) {$r_+$};
\draw (0,8)--(2,10)node [midway, below, sloped] (TextNode) {$r_-$};
\draw (2,10)--(4,8)node [midway, below, sloped] (TextNode) {$r_-$};
\draw (2,10)--(0,12);
\draw (2,10)--(4,12);
\draw [line width=2pt] (0,8) -- (0,12) node [midway, above, sloped] (TextNode) {$r=0$};
\draw [line width=2pt] (4,12) -- (4,8) node [midway, above, sloped] (TextNode) {$r=0$};
\draw [line width=2pt] (8,0) -- (8,4) node [midway, above, sloped] (TextNode) {$r=0$};
\draw[line width=2pt] (4,4)--(8,4)node [midway, below, sloped] (TextNode) {$r=\infty$};
\draw (6,6)--(8,4);
\draw (6,6)--(8,8);
\draw[line width=2pt] (4,8)--(8,8)node [midway, above, sloped] (TextNode) {$r=\infty$};
\draw [line width=2pt](8,8)--(8,12) node [midway, above, sloped] (TextNode) {$r=0$};
\draw (8,4)--(10,6);
\draw (8,8)--(10,6);
\draw (8,8)--(10,10);
\draw (8,12)--(10,10);
\draw (8,0)--(10,2);
\draw (8,4)--(10,2);
\node at (2,0.5){$\rII_+$};
\node at (6,7){$\rIV_-$};
\node at (6,5){$\rIV_+$};
\node at (3,2){$\rIII_-$};
\node at (3,10){$\rIII_-$};
\node at (1,2){$\rIII_+$};
\node at (1,10){$\rIII_+$};
\node at (2,4){$\rII_-$};
\node at (0,6){$\rI_-$};
\node at (4,6){$\rI_+$};
\node at (2,8){$\rII_+$};
\node at (2,11.5){$\rII_-$};
\node at (9.5,4){$\rII_-$};
\node at (9.5,0.5){$\rII_+$};
\node at (9.5,8){$\rII_+$};
\node at (9.5,11.5){$\rII_-$};
\node at (8,6){$\rI_-$};
\node at (9,10){$\rIII_+$};
\node at (9,2){$\rIII_+$};
\end{tikzpicture}
\caption{Extension of the RNdS spacetime. The thick, vertical lines indicate the curvature singularities and the thick, horizontal lines conformal infinity.  The spacetime can be infinitely continued to the top, bottom, left and right.}
\label{fig:RNdS}
\end{figure}
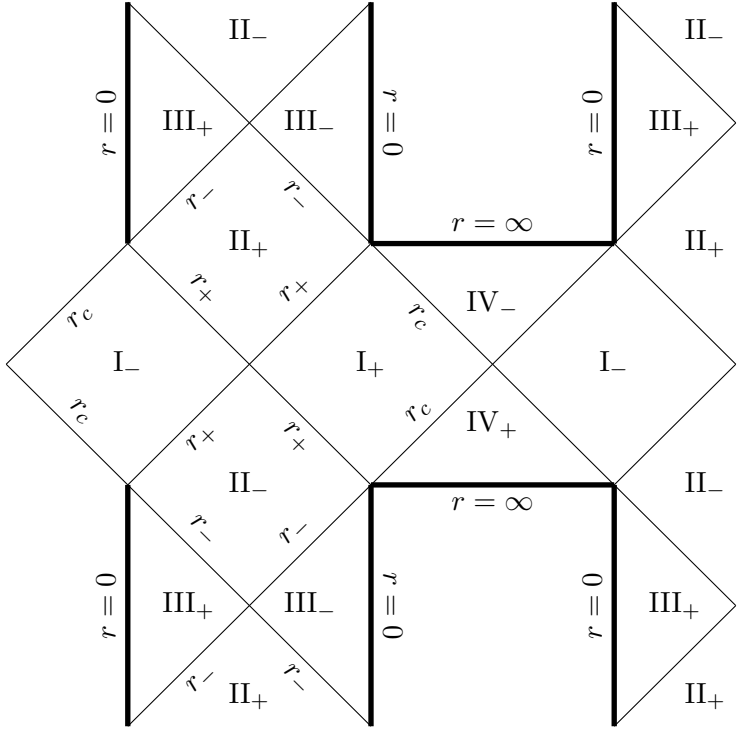

A conformal diagram of the maximal extension of the generic black hole case of RNdS, \cf \cite{BrillHayward}, is shown in Fig.~\ref{fig:RNdS}. This diagram can be extended infinitely in all directions by continuing the pattern. From the above discussion on the sign of $f_+(r)$ it follows that $\del_t$ is time-like in regions $\rI_\pm$ and $\rIII_\pm$ but space-like in regions $\rII_\pm$ and $\rIV_\pm$. We may think of region $\rI_+$ as the region outside of but still in causal contact with the black hole, represented by region $\rII_+$ to the upper left of it. An observer which entered region $\rIV_-$ can no longer fall into the black hole, but will continue to ``conformal infinity'' $r = \infty$.

\begin{figure}
\centering
\begin{tikzpicture}[scale=0.8]
\draw (0,0)--(4,4);
\draw[line width=2pt] (4,4)--(4,0)node[midway,above, sloped] (TextNode) {$r=0$};
\draw [line width=2pt](0,0)--(4,0)node[midway,below, sloped] (TextNode) {$r=\infty$};
\draw[line width=2pt] (0,0)--(0,4)node[midway,above, sloped] (TextNode) {$r=0$};
\draw (0,4)--(4,0);
\draw (0,0)--(2,2)node[midway,above, sloped] (TextNode) {$r_c$};
\draw (2,2)--(4,4)node[midway,above, sloped] (TextNode) {$r_c$};
\draw (0,4)--(2,2)node[midway,above, sloped] (TextNode) {$r_c$};
\draw (2,2)--(4,0)node[midway,above, sloped] (TextNode) {$r_c$};
\draw [line width=2pt](0,4)--(4,4)node[midway,above, sloped] (TextNode) {$r=\infty$};
\node at (2,1){$\rIV_+$};
\node at (2,3){$\rIV_-$};
\node at (1,2){$\rI_+$};
\node at (3,2){$\rI_-$};
\end{tikzpicture}
\caption{The generic naked singularity case of RNdS.}
\label{fig:NakedSingularity}
\end{figure}
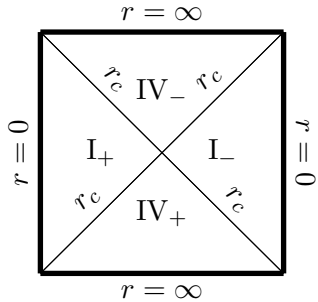

As we will also briefly discuss the naked singularity case at the end, we also note that its conformal diagram, shown in Fig.~\ref{fig:NakedSingularity}, has the same form as that of dS, Fig~\ref{fig:dS}, with the origin of coordinates $r = 0$ replaced by a singularity and the Hubble horizon $r_H$ replaced by the cosmological horizon. The singularity is now visible from region $\rI_+$, i.e., it is not hidden behind a horizon.

According to the thin shell formalism \cite{Israel1966, Poisson}, the metrics induced on the shell by the two metrics $g_\pm$ must coincide. From the angular component of the metrics \eqref{eq:metrics}, it follows that the shell must be located at the same value of the radial coordinate \wrt the two patches. We denote its value by $R$. The situation we have in mind is that in the dS region, $r$ ranges from $0$ to $R$, and in the RNdS region from $R$ to $\infty$. However, we will also consider other situations (in particular, the situation where $r$ decreases from $R$ to $0$ in the RNdS region will naturally occur). Furthermore, again by the condition that the induced metrics coincide, the proper time $\tau$ of the shell is invariantly defined (up to an additive constant). The four velocity of the shell \wrt proper time is thus
\beq
 u_\pm^\mu = ( u^0_\pm, \dot R, 0, 0 )
\eeq
in the above coordinate system (the derivative \wrt $\tau$ is denoted by a dot). Notice that the value of $u^0$ differs in the two regions separated by the shell. Using the normalization of the four-velocity, we have
\beq
\label{eq:u_0_2}
 (u^0_\pm)^2 = f_\pm(R)^{-2} \left( \dot R^2 + f_\pm(R) \right).
\eeq
The normalized normal of the shell is, up to a sign, given by
\beq
 n_{\pm \mu} = ( - \dot R, u^0_\pm, 0, 0).
\eeq
The sign is by convention chosen such that $n_-^\mu$ points outward \wrt the dS region and $n_+^\mu$ points inward \wrt the RNdS region, i.e., they both point into the RNdS region. We thus have
\beq
\label{eq:n1_pm}
 n_\pm^1 = \eps_\pm \sqrt{\dot R^2 + f_\pm(R)},
\eeq
with $\eps_\pm \in \{ +1, -1 \}$ indicating the direction of the radial component of the normal. In the ``standard'' situation described above, i.e., $r$ ranging from $0$ to $R$ in the dS region and from $R$ to $\infty$ in the RNdS region, we would have $\eps_- = \eps_+ = 1$. If, for example, the radial coordinate decreases from the shell in the RNdS region, we would have $\eps_+ = -1$.

For a spherically symmetric shell of pressureless dust, the surface stress tensor of the shell is of the form
\beq
 S_{ab} = \sigma u_a u_b,
\eeq
with $a$, $b$ referring to coordinates intrinsic to the shell and $\sigma$ independent of the angular coordinates. From the vanishing divergence of $S_{ab}$, it follows that
\beq
 M = 4 \pi R^2 \sigma = \const ,
\eeq
which is interpreted as the rest mass of the shell. We will mostly assume it to be positive, but will also briefly consider the case of negative $M$. The surface stress tensor $S_{ab}$ is related to the difference of extrinsic curvature $K_{a b}$ of the shell \wrt $g_+$ and $g_-$ via the Lanczos equation \cite{Lanczos1924, Israel1966, Poisson}
\beq
 8 \pi S_{ab} = - K_{+ ab} + K_+ h_{ab} + K_{- ab} - K_- h_{ab},
\eeq
with $h_{ab}$ the induced metric of the shell and $K = K_{ab} h^{ab}$. Contracting with $u^a u^b$ on both sides yields \cite{Boulware1973, Poisson} 
\beq
 8 \pi \sigma = - \frac{2}{R} (n^1_+ - n^1_-),
\eeq
or
\beq
\label{eq:M}
 M = - R (n^1_+ - n^1_-). 
\eeq

\section{Shell dynamics}

With \eqref{eq:n1_pm} and \eqref{eq:M} we already have the fundamental equations governing the dynamics of the shell. From these, one finds that
\beq
 \eps_\pm = \sign(M) \sign (2 m R - Q^2 \mp M^2 ).
\eeq
We see that a sign change of $\eps_\pm$ occurs at
\beq
 R_\pm = \frac{Q^2 \pm M^2}{2 m}.
\eeq
In the case $M > 0$, to which we now restrict for the time being, this means that $\eps_\pm = 1$ for $R > R_\pm$ and $\eps_\pm = - 1$ for $R < R_\pm$. From the assumed positivity of $m$ it follows that $R_- < 0$ for $M^2 > Q^2$, so that $\eps_- = 1$ in that case, irrespective of the value of $R$.
On the other hand, $R_+$ is always positive, so that there is always a range of the shell radius $R$ such that $\eps_+ = -1$. Another direct consequence of \eqref{eq:n1_pm} and \eqref{eq:M} is
\beq
 m = \eps_- M \sqrt{1 - \kappa^2 R^2 + \dot R^2} - \frac{M^2 - Q^2}{2 R},
\eeq
which expresses the gravitational mass as a sum of a ``kinetic'' energy and a ``potential'' energy. Note that the ``kinetic'' term also depends on the cosmological constant, and that the ``potential'' energy differs from the naive one by a factor $1/2$, due to the fact that ``only half of the shell is subject to its own potential'', see also \cite{Boulware1973}. Finally, we can use \eqref{eq:n1_pm} and \eqref{eq:M} to determine a dynamical equation for $R$,
\beq
\label{eq:dotR}
 \dot R^2 = V(R),
\eeq
with
\beq
\label{eq:V}
 V(R) = \frac{1}{M^2 R^2} \left( \kappa^2 M^2 R^4 + (m^2 - M^2) R^2 + m (M^2 - Q^2) R + \frac{1}{4} (M^2 - Q^2)^2 \right).
\eeq
This is consistent with the result of \cite{Boulware1973} and \cite{YamanakaEtAl} in the limits $\kappa \to 0$ and $Q \to 0$, respectively.

We will not try to solve \eqref{eq:dotR} explicitly. Instead, we are interested in qualitative features of the trajectories, which are governed by the turning points of $R$, where $V(R) = 0$. First of all, we notice that if $R_0$ is a simple root of $V(R)$, then a trajectory $R(\tau)$ approaching it will reach it in finite proper time $\tau$. Explicit expressions for the roots of $V(R)$ can be given, but they are quite lengthy and not illuminating. However, the qualitative behaviour of the roots can be discussed in elementary terms. To begin with, there are at most two roots in the range $R>0$: By the absence of a cubic term in the expression in brackets in \eqref{eq:V}, the sum of the four roots vanishes (degenerate roots are counted according to their multiplicity), so there can be at most three positive roots. Furthermore, $V(R) \to + \infty$ for $R \to + \infty$ and $V(0) \geq 0$, which, together with the previous statement, implies that there are at most two positive roots. In the case $M^2 \neq Q^2$, for which $V(R)$ is strictly positive for $R \to 0$, we can even conclude that there are either no or two positive roots (again counted according to multiplicity). We call the greater of the two positive roots the outer turning point and the lesser of the two the inner turning point. Between the two turning points, $V(R)$ is negative, so that \eqref{eq:dotR} has no solution. Hence, between the turning points lies a forbidden region that the shell radius can not enter. 

\begin{figure}
\centering
\includegraphics{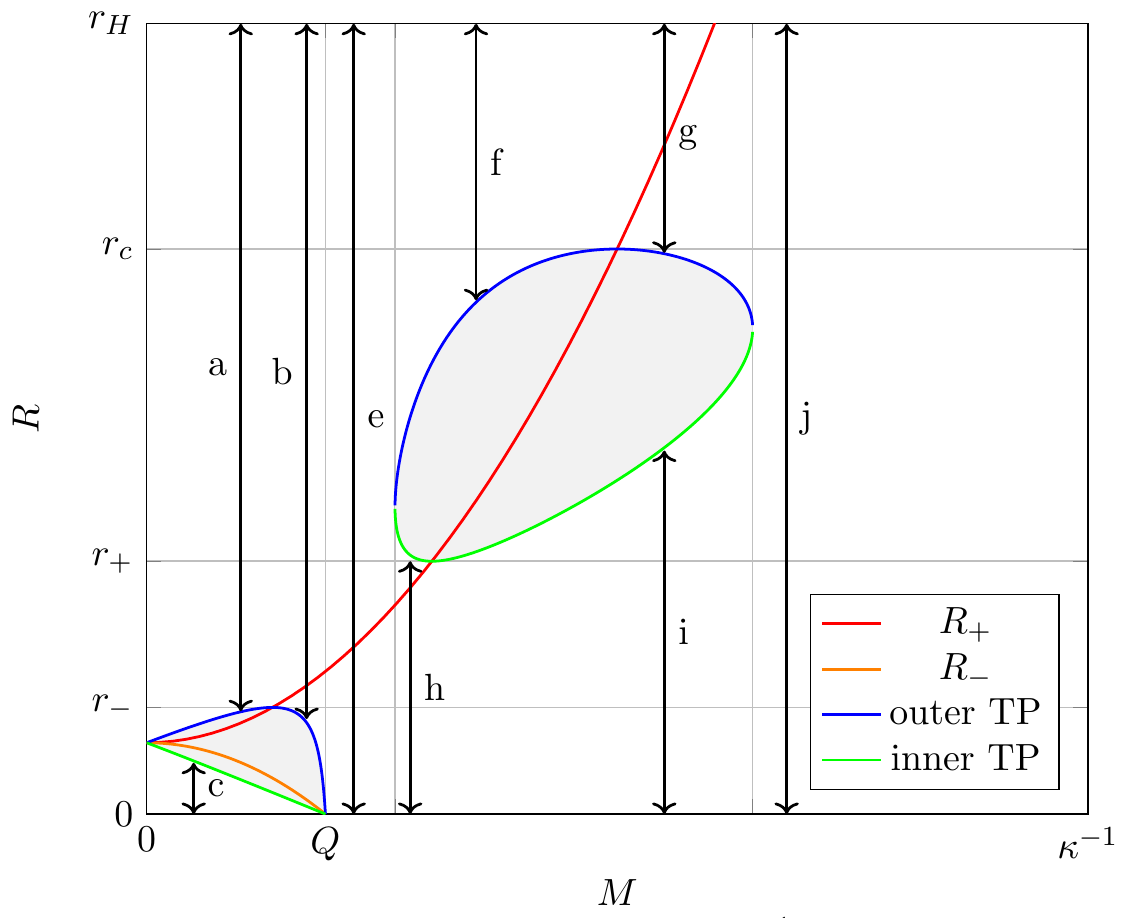}
\caption{\label{fig:Q<m}The position of the outer and inner turning points (in blue and green) for $m = 0.2 \kappa^{-1}$, $Q = 0.19 \kappa^{-1}$ as a function of $M$. Also shown (in red and orange) are the values $R_\pm$ at which $\eps_\pm$ changes its sign. The forbidden regions are shaded in gray. The black arrows indicate the possible trajectories of the shell.}
\end{figure}

\begin{figure}
\centering
\includegraphics{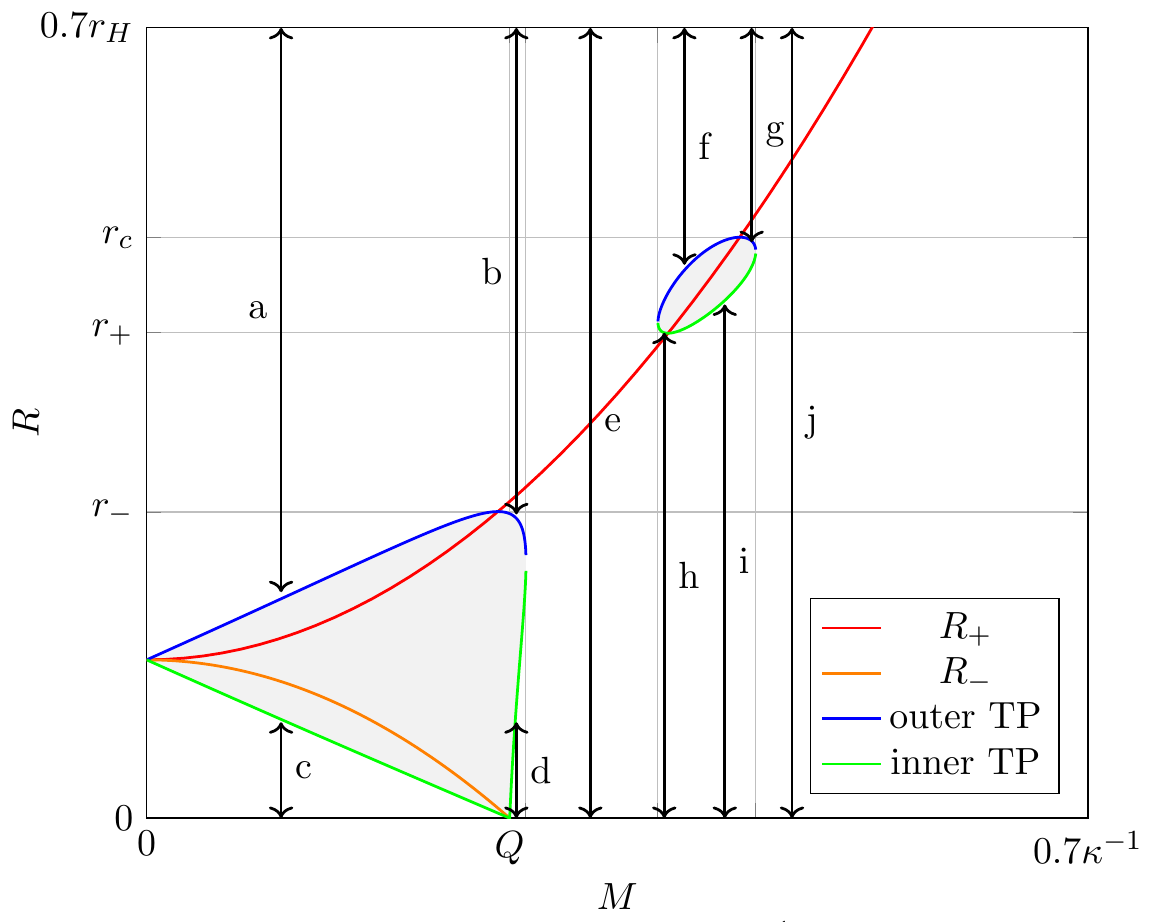}
\caption{\label{fig:Q>m}Same as Fig.~\ref{fig:Q<m} but for the near extremal case $Q > m$ ($m = 0.26 \kappa^{-1}$, $Q = 0.27 \kappa^{-1}$).}
\end{figure}

In Fig.~\ref{fig:Q<m} the two turning points are plotted as a function of $M$ for fixed values of $m$ and $Q$ (an analogous diagram for the case $Q = 0$ can be found in \cite{YamanakaEtAl}). We see that there are two finite ranges of $M$ for which turning points are present. For small values of $M$, both turning points lie in region $\rIII$, while for greater values of $M$ both turning points lie in region $\rI$. In between and for yet larger values of $M$, no turning points exist. While Fig.~\ref{fig:Q<m} only shows the turning points for fixed specific values of $m$ and $Q$, this behavior is generic, as will be proven below. However, apart from the special cases of extreme black holes, there is one case distinction that should be made: For $M^2 = Q^2$, one obviously has no positive root of $V(R)$ for $m \geq Q$ and a single positive root for $m < Q$. Hence, for the study of the turning points in the parameter region $M^2 \simeq Q^2$, it is appropriate to distinguish the two cases. While Fig.~\ref{fig:Q<m} shows the turning points in the first case, the turning points for $Q > m$, called the  near extremal case in the following, are shown in Fig.~\ref{fig:Q>m}.

From the above discussion, we can already identify three qualitatively different evolutions (a finer distinction will be made below): In the absence of positive roots of $V(R)$, i.e., turning points of $R$, the shell radius will decrease monotonically until it hits the singularity at $R = 0$ (of course there is also the time reverse of the process, a shell emerging from the singularity and expanding to $R \to \infty$). In Fig.~\ref{fig:Q<m} and Fig~\ref{fig:Q>m} this is the case for the trajectories labelled e) and j). In the presence of two turning points, we have two types of solutions: One is expanding from the singularity at $R = 0$, reaching the inner turning point, and then recollapsing into the singularity. In Fig.~\ref{fig:Q<m} and Fig.~\ref{fig:Q>m} this is the case for the trajectories labelled c), h), and i) and in Fig.~\ref{fig:Q>m} also for the trajectory labelled d). The other possibility is to start at $R = \infty$, collapse until the outer turning point is reached, and then reexpand again to $R \to \infty$. In Fig.~\ref{fig:Q<m} and Fig.~\ref{fig:Q>m} this is the case for the trajectories labelled a), b), f), and g).

We would like to prove that the behaviour of the turning points shown in Figs.~\ref{fig:Q<m} and \ref{fig:Q>m} is generic.
From \eqref{eq:u_0_2}, we conclude that 
a turning point at $R$ is only possible if $f_+(R) \geq 0$ and $f_-(R) \geq 0$. The first condition excludes turning points in regions $\rII$ and $\rIV$ of RNdS, compatible with Figs.~\ref{fig:Q<m} and \ref{fig:Q>m}, and the second condition excludes turning points in region $\tilde \rII$ of dS. 
Furthermore, as we learned above that $V(R)$ must be negative between two positive turning points, we can also conclude that the two turning points must lie in the same region of RNdS, as otherwise they would be separated by region $\rII$, where $V(R)$ is positive, by the following equivalent form of $V(R)$:
\beq
 V(R) = \frac{R^2}{4 M^2} \left( f_-(R) - f_+(R) - \frac{M^2}{R^2} \right)^2 - f_+(R).
\eeq
Again, this feature is compatible with the behaviour shown in Figs.~\ref{fig:Q<m} and \ref{fig:Q>m}.

To fully characterize the qualitative behaviour of the shell trajectories, we have to consider the signs $\eps_\pm$ of the radial component of the normal vector at the turning point. From \eqref{eq:n1_pm}, we see that a sign change of $\eps_\pm$ can only occur where $f_\pm(R) < 0$, i.e., in regions $\rII$ or $\rIV$ for $\eps_+$ and in region $\tilde \rII$ for $\eps_-$. For a trajectory reflected at an outer turning point $R$, we can distinguish between the cases $R \geq R_+$, in which case $\eps_+$ is positive throughout, and $R < R_+$, in which case $\eps_+$ changes signs twice, once before and once after the reflection of the trajectory. In Figs.~\ref{fig:Q<m} and \ref{fig:Q>m} the trajectories a) and f) correspond to the first and the trajectories b) and g) to the second case. 

As for $\eps_-$, we recall that $R_- \leq 0$ for $M^2 \geq Q^2$, so that in this case the sign change never happens and $\eps_- = 1$ throughout the trajectory. For the case $M^2 < Q^2$, we will show below that, as in Figs.~\ref{fig:Q<m} and \ref{fig:Q>m}, there are always two turning points, with $R_-$ in between, i.e., in the forbidden region. Hence, also in this case, a sign change of $\eps_-$ does not occur. However, for a trajectory starting at $R = 0$, $\eps_- = -1$ in the range $M^2 < Q^2$, so such trajectories have $\eps_- = -1$ throughout.

To learn more about the relation between turning points and $R_\pm$, one easily verifies the relation
\beq
\label{eq:V_f}
 V(R_\pm) = - f_\pm(R_\pm).
\eeq
Hence, for $R_+$ in region $\rI$ or $\rIII$, we see that $V(R_+) < 0$, i.e., $R_+$ lies in the forbidden region between two turning points. At the boundaries of these regions, i.e., for $R_+ = r_{-/+/c}$, we also have a root of $V(R)$, i.e., a turning point. Hence, $R_+$ crosses a turning point when passing through any of the RNdS horizons, a feature that we also identify in Figs.~\ref{fig:Q<m} and \ref{fig:Q>m}. As for $R_-$, we first notice that for $M$ near $Q$ (for $M < Q$), we have $R_-$ close to $0$, i.e., both in $\tilde \rI$ and $\rIII$, so that by \eqref{eq:V_f} $R_-$ is in the forbidden region. It must thus be contained between two turning points, which, by the argument given above, both lie in $\rIII$. If we now let $M$ decrease, $R_+$ decreases and $R_-$ increases monotonically, until they both coincide with $\frac{Q^2}{2m}$ for $M \to 0$. It follows that for $M < Q$, there are always two turning points in $\rIII$, and the inner one converges to $0$ for $M \to Q$.
All these features are present in Figs.~\ref{fig:Q<m} and \ref{fig:Q>m}, the only distinction between the two being that in the first case also the outer turning point goes to $0$ as $M \to Q$, while in the latter case the outer turning point is positive for $M = Q$.
Furthermore, as $R_+$ starts in $\rIII$ for $M = 0$ and then passes through $\rII$, $\rI$, and $\rIV$ as $M$ increases and is contained in the forbidden region between two turning points when passing through $\rIII$ or $\rI$, this proves that for all values of $m$ and $Q$ there will be a range of values of $M$ for which two turning points in $\rIII$ exist and a range for which two turning points in $\rI$ exist (as is the case in Fig.~\ref{fig:Q<m} and Fig.~\ref{fig:Q>m} for specific values of $m$ and $Q$).

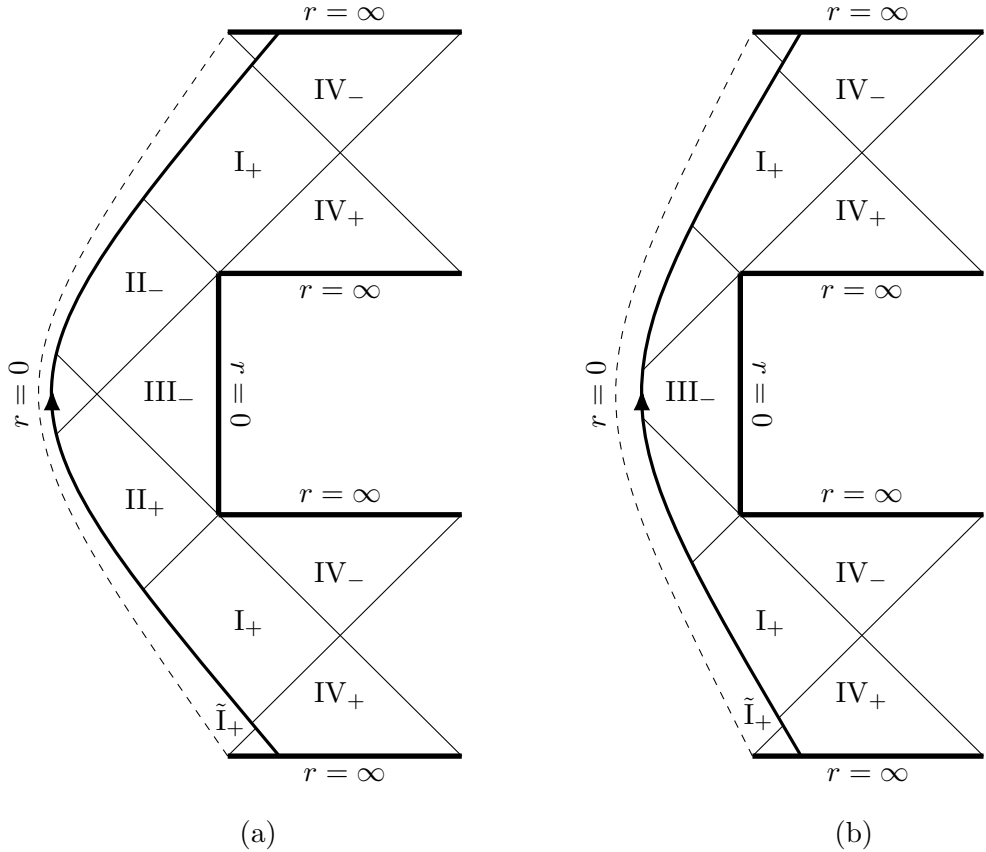
\begin{figure}
\centering
\begin{minipage}{.47\textwidth}
\begin{tikzpicture}[scale=0.8]
\draw[line width=2pt] (4.15,0)--(8,0)node[midway,below, sloped] (TextNode) {$r=\infty$};
\draw[line width=2pt] (4,4)--(8,4)node[midway,above, sloped] (TextNode) {$r=\infty$};
\draw[line width=2pt] (4,8)--(4,4)node[midway,above, sloped] (TextNode) {$r=0$};
\draw[line width=2pt] (4,8)--(8,8)node[midway,below, sloped] (TextNode) {$r=\infty$};
\draw[line width=2pt] (4.15,12)--(8,12)node[midway,above, sloped] (TextNode) {$r=\infty$};
\draw (6,2)--(8,0);
\draw (6,2)--(8,4);
\draw (6,2)--(4,4);
\draw (6,10)--(8,8);
\draw (6,10)--(8,12);
\draw (6,10)--(4,8);
\draw[very thick] (5,0)..controls(0,6)..(5,12);
\draw[dashed](4.15,0)..controls(0,6)..(4.15,12)node[midway,above, sloped] (TextNode) {$r=0$};
\draw[->,>=Latex, very thick](1.26,6)--(1.26,6.1);
\draw (2,6)--(4,8);
\draw (2,6)--(4,4);
\draw (6,10)--(4.55,11.45);
\draw (6,2)--(4.55,0.55);
\draw (4.6,0.45)--(4.15,0);
\draw (4.6,11.55)--(4.15,12);
\draw (4,4)--(2.75,2.75);
\draw (4,8)--(2.75,9.25);
\draw (2,6)--(1.32,5.32);
\draw (2,6)--(1.32,6.68);
\node at (4.5,2.2){$\rI_+$};
\node at (4.5,9.8){$\rI_+$};
\node at (6,1){$\rIV_+$};
\node at (6,3){$\rIV_-$};
\node at (6,9){$\rIV_+$};
\node at (6,11){$\rIV_-$};
\node at (3.2,6){$\rIII_-$};
\node at (2.8,4.2){$\rII_+$};
\node at (2.8,7.8){$\rII_-$};
\node at (4.2,0.6){$\tilde \rI_+$};
\end{tikzpicture}
\captionof*{figure}{(a)}
\end{minipage}
\hspace{.1cm}
\begin{minipage}{.47\textwidth}
\begin{tikzpicture}[scale=0.8]
\draw[line width=2pt] (4.2,0)--(8,0)node[midway,below, sloped] (TextNode) {$r=\infty$};
\draw[line width=2pt] (4,4)--(8,4)node[midway,above, sloped] (TextNode) {$r=\infty$};
\draw[line width=2pt] (4,8)--(4,4)node[midway,above, sloped] (TextNode) {$r=0$};
\draw[line width=2pt] (4,8)--(8,8)node[midway,below, sloped] (TextNode) {$r=\infty$};
\draw[line width=2pt] (4.2,12)--(8,12)node[midway,above, sloped] (TextNode) {$r=\infty$};
\draw (6,2)--(8,0);
\draw (6,2)--(8,4);
\draw (6,2)--(4,4);
\draw (6,10)--(8,8);
\draw (6,10)--(8,12);
\draw (6,10)--(4,8);
\draw[very thick] (5,0)..controls(1.5,6)..(5,12);
\draw[dashed] (4.2,0)..controls(1.2,6)..(4.2,12)node[midway,above, sloped] (TextNode) {$r=0$};
\draw[->,>=Latex, very thick](2.38,6)--(2.38,6.1);
\draw (6,2)--(4.64,0.64);
\draw (4,4)--(3.2,3.2);
\draw (4,4)--(2.4,5.6);
\draw (6,10)--(4.64,11.36);
\draw (4,8)--(3.2,8.8);
\draw (4,8)--(2.4,6.4);
\draw (4.7,0.5)--(4.2,0);
\draw (4.7,11.5)--(4.2,12);
\node at (4.5,2.2){$\rI_+$};
\node at (4.5,9.8){$\rI_+$};
\node at (6,1){$\rIV_+$};
\node at (6,3){$\rIV_-$};
\node at (6,9){$\rIV_+$};
\node at (6,11){$\rIV_-$};
\node at (3.2,6){$\rIII_-$};
\node at (4.3,0.7){$\tilde \rI_+$};
\end{tikzpicture}
\captionof*{figure}{(b)}
\end{minipage}
\caption{\label{fig:a_b}Paths a) and b) describing a collapsing shell with a turning point inside the Cauchy horizon. The turning point lies in the region $\rIII_+$ or $\rIII_-$ depending on the sign of $\eps_+$ at the turning point. The sign of $\eps_-$ is positive in both cases.}
\end{figure}

Having now understood the qualitative features of the turning points in the generic black hole case, we discuss the shell trajectories through space-time diagrams. Trajectories of type a) and b) are shown in Fig.~\ref{fig:a_b}. They are distinguished by $\eps_+$ at the turning point, which is $+1$ for a) and $-1$ for b). It follows that in the first case, the turning point lies in $\rIII_+$ (where $r$ increases towards the right), whereas it lies in $\rIII_-$ for case b). In both cases, the shell does not collapse into the singularity, but emerges into another asymptotically de Sitter region, where it expands to $R \to \infty$. Similar trajectories were also found for a vanishing cosmological constant, \cf case 2c) of \cite{Boulware1973}.

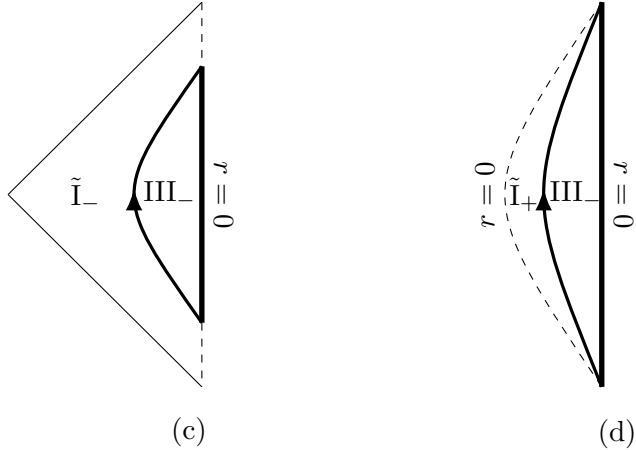
\begin{figure}
\centering
\begin{minipage}{.3\textwidth}
\begin{tikzpicture}[scale=0.85]
\draw [line width=2pt](6,5)--(6,1)node[midway,above, sloped] (TextNode) {$r=0$};
\draw[very thick] (6,1)..controls (4.6,3)..(6,5);
\draw[->,>=Latex, very thick](4.95,3)--(4.95,3.1);
\draw[dashed] (6,6)--(6,5);
\draw[dashed] (6,1)--(6,0);
\draw (3,3)--(6,0);
\draw (3,3)--(6,6);
\node at (4.2,3){$\tilde \rI_-$};
\node at (5.5,3){$\rIII_-$};
\end{tikzpicture}
\captionof*{figure}{(c)}
\end{minipage}
\hspace{1cm}
\begin{minipage}{.25\textwidth}
\begin{tikzpicture}[scale=0.85]
\draw [line width=2pt](2,6)--(2,0)node[midway,above, sloped] (TextNode) {$r=0$};
\draw[very thick] (2,0)..controls (0.8,3)..(2,6);
\draw[->,>=Latex, very thick](1.1,3)--(1.1,3.1);
\draw[dashed] (2,0)..controls (0,3)..(2,6)node[midway,above, sloped] (TextNode) {$r=0$};
\node at (0.8,3){$\tilde \rI_+$};
\node at (1.6,3){$\rIII_-$};
\end{tikzpicture}
\captionof*{figure}{(d)}
\end{minipage}
\caption{\label{fig:c_d}The path c) corresponds to the bubble trajectory. The shell expands in the de Sitter space region $\tilde \rI_-$ and collapses after reaching the inner turning point. The inside is the region $\rIII_-$ of the RNdS space. For $m> Q$, also a trajectory of type d) is possible.}
\end{figure}

The paths c) and d) are sketched in Fig.~\ref{fig:c_d}, the latter being possible only in the near extremal case $m < Q$, \cf Fig.~\ref{fig:Q>m}. In both cases, $\eps_+ = -1$, corresponding to a turning point in region $\rIII_-$. In the first case, $\eps_- = -1$, corresponding to a turning point in $\tilde \rI_-$, while for case d) $\eps_- = +1$, so that the turning point lies in $\tilde \rI_+$. As no horizons are crossed, the whole trajectory lies in these regions. Trajectory c) can be interpreted as the spontaneous creation, in de Sitter space, of a charged shell and an oppositely charged singularity. The shell expands for a while but then recollapses into the singularity. Such a ``bubble'' trajectory is also possible in the asymptotically flat case (it does not appear in \cite{Boulware1973} only because the case $\eps_- = -1$ was excluded from the discussion). Trajectory d) can be interpreted as the spontaneous creation of parts of dS and RNdS space, both ranging in $r$ from $0$ to $R$, and glued together at $R$. The shell radius $R$ expands from $0$ to some maximum, and then recollapses to $0$, thereby terminating the spacetime. Trajectories of this type are not possible for vanishing cosmological constant \cite{Boulware1973} or vanishing charge \cite{YamanakaEtAl}.

Path e) is a trajectory without turning point, sketched in Fig.~\ref{fig:e}. We have $\eps_+ = +1$ for $R > R_+$, the latter being in region $\rII$. Hence, the trajectory passed through regions $\rIV_+$ and $\rI_+$ before. After traversing $\rII_+$, $\eps_+ = -1$, so that the shell then traverses $\rIII_-$ to finally collide into the singularity $r = 0$, which formed outside of the shell. The time reverse of the process describes a ``shell explosion'' leading to a charged shell in de Sitter space, expanding to $R = \infty$. A trajectory of this type is also possible for a vanishing cosmological constant, \cf case 1) of \cite{Boulware1973}.

\begin{figure}
\centering
\begin{minipage}{.48\textwidth}
\begin{tikzpicture}[scale=0.85]
\draw[line width=2pt] (4.1,0)--(8,0)node[midway,below, sloped] (TextNode) {$r=\infty$};
\draw[line width=2pt] (4,4)--(8,4)node[midway,above, sloped] (TextNode) {$r=\infty$};
\draw[line width=2pt] (4,7.5)--(4,4)node[midway,above, sloped] (TextNode) {$r=0$};
\draw (6,2)--(8,0);
\draw (6,2)--(8,4);
\draw (6,2)--(4,4);
\draw[very thick] (5,0)..controls (2,4)..(4,7.5);
\draw[dashed] (4.1,0)..controls(1.2,4.1)..(4,7.5) node[midway,below, sloped] (TextNode) {$r=0$};
\draw (6,2)--(4.6,0.6);
\draw (4,4)--(2.95,2.95);
\draw (4,4)--(2.8,5.2);
\draw (4.6,0.5)--(4.1,0);
\draw[->,>=Latex, very thick](2.62,4)--(2.62,4.1);
\node at(6,1){$\rIV_+$};
\node at(6,3){$\rIV_-$};
\node at(4.3,2.2){$\rI_+$};
\node at(3.3,4){$\rII_+$};
\node at(3.5,5.2){$\rIII_-$};
\node at (3.8,1.1){$\tilde \rI_+$};
\end{tikzpicture}
\end{minipage}
\begin{minipage}{.48\textwidth}
\begin{tikzpicture}[scale=0.85]
\draw[line width=2pt] (4.15,8)--(8,8)node[midway,above, sloped] (TextNode) {$r=\infty$};
\draw[line width=2pt] (4,4)--(8,4)node[midway,below, sloped] (TextNode) {$r=\infty$};
\draw[line width=2pt] (4,4)--(4,1)node[midway,above, sloped] (TextNode) {$r=0$};
\draw[very thick](4,1)..controls(2,4)..(5,8);
\draw[dashed](4,1)..controls(1.2,4)..(4.15,8)node[midway,above, sloped] (TextNode) {$r=0$};
\draw[->,>=Latex, very thick](2.62,4)--(2.62,4.1);
\draw (6,6)--(8,8);
\draw (6,6)--(8,4);
\draw (6,6)--(4,4);
\draw (4,4)--(2.95,5.05);
\draw (4,4)--(2.85,2.85);
\draw (6,6)--(4.55,7.45);
\draw (4.65,7.5)--(4.15,8);
\node at(6,5){$\rIV_+$};
\node at(6,7){$\rIV_-$};
\node at(4.5,5.7){$\rI_+$};
\node at(3.3,4){$\rII_-$};
\node at(3.5,2.8){$\rIII_-$};
\node at (3.8,6.9){$\tilde \rI_+$};
\end{tikzpicture}
\end{minipage}
\captionof*{figure}{(e)}
\caption{\label{fig:e}Path e) may describe a collapse (left) or an explosion (right) depending on the sign of $\dot{R}$. $\eps_+$ changes sign in regions of type $\rII$, thus the world line must evolve through region $\rI_+$.}
\end{figure}

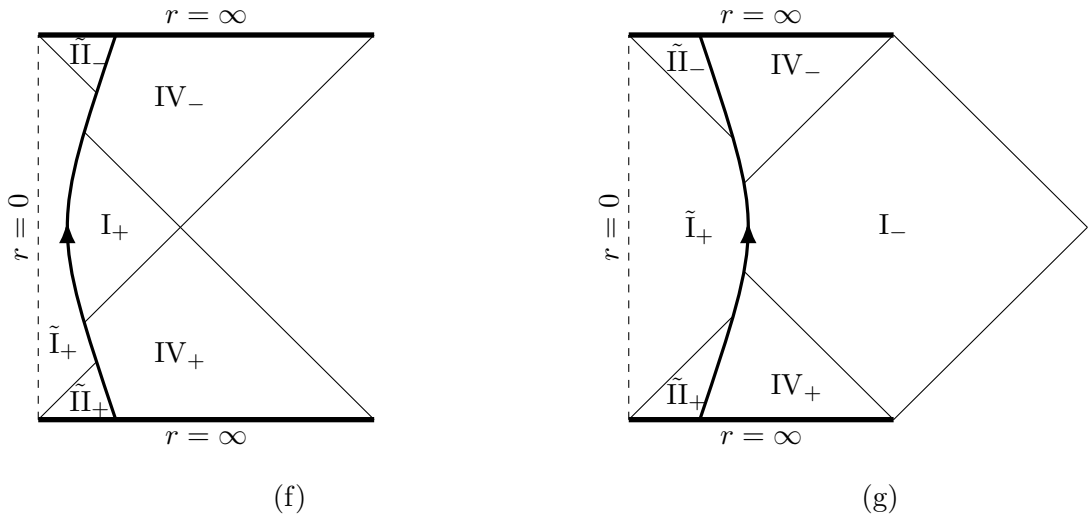
\begin{figure}
\centering
\begin{minipage}{.48\textwidth}
\begin{tikzpicture}[scale=.85]
\draw [line width=2pt](0.8,0)--(6,0)node[midway,below, sloped] (TextNode) {$r=\infty$};
\draw [line width=2pt](0.8,6)--(6,6)node[midway,above, sloped] (TextNode) {$r=\infty$};
\draw[very thick] (2,0).. controls (1,3)..(2,6);
\draw (3,3)--(6,6);
\draw[->,>=Latex, very thick](1.25,3)--(1.25,3.1);
\draw (3,3)--(6,0);
\draw (1.5,1.5)--(3,3);
\draw (1.5,4.5)--(3,3);
\draw (0.8,6)--(1.7,5.1);
\draw (0.8,0)--(1.7,0.9);
\draw[dashed] (0.8,0)--(0.8,6)node[midway,above, sloped] (TextNode) {$r=0$};
\node at (3,5){$\rIV_-$};
\node at (3,1){$\rIV_+$};
\node at (2,3){$\rI_+$};
\node at (1.2,1.2){$\tilde \rI_+$};
\node at (1.6,0.3){$\tilde \rII_+$};
\node at (1.6,5.7){$\tilde \rII_-$};
\end{tikzpicture}
\captionof*{figure}{(f)}
\end{minipage}
\begin{minipage}{.48\textwidth}
\begin{tikzpicture}[scale=.85]
\draw [line width=2pt](1.9,0)--(6,0)node[midway,below, sloped] (TextNode) {$r=\infty$};
\draw [line width=2pt](1.9,6)--(6,6)node[midway,above, sloped] (TextNode) {$r=\infty$};
\draw[very thick] (3,0).. controls (4,3)..(3,6);
\draw[->,>=Latex, very thick](3.75,3)--(3.75,3.1);
\draw (3.7,2.3)--(6,0);
\draw (3.7,3.7)--(6,6);
\draw (6,0)--(9,3);
\draw (6,6)--(9,3);
\draw (1.9,0)--(3.5,1.6);
\draw (1.9,6)--(3.5,4.4);
\draw[dashed] (1.9,0)--(1.9,6)node[midway,above, sloped] (TextNode) {$r=0$};
\node at (3,3){$\tilde \rI_+$};
\node at (2.8,0.4){$\tilde \rII_+$};
\node at (2.8,5.6){$\tilde \rII_-$};
\node at (6,3){$\rI_-$};
\node at (4.5,0.5){$\rIV_+$};
\node at (4.5,5.5){$\rIV_-$};
\end{tikzpicture}
\captionof*{figure}{(g)}
\end{minipage}
\caption{\label{fig:f_g}For the trajectories f) and g), the turning points lie in regions of type $\rI$. The collapsing shell reaches the outer turning point in region $\rI_+$ (path f)) or $\rI_-$ (path g)) depending on the sign of $\eps_+$ at the turning point.}
\end{figure}

Trajectories f) and g) are sketched in Fig.~\ref{fig:f_g}. In the first case, $\eps_+$ is always positive, so the shell passes through region $\rI_+$, while in the second case, $\eps_- = -1$ at the turning point, so that the shell passes through region $\rI_-$. Both trajectories describe a shell that contracts from infinite extension up to a minimal radius and then expands again to $R = \infty$. The difference is that for trajectory g) the radius $r$ decreases in the RNdS region away from the shell at the turning point. Trajectories analogous to f) and g) do not exist for vanishing cosmological constant \cite{Boulware1973}. They do, however, also exist in the uncharged case $Q = 0$ \cite{YamanakaEtAl}. Hence, we can interpret the bounce as due to the cosmological constant, not the repulsion due to the charge.

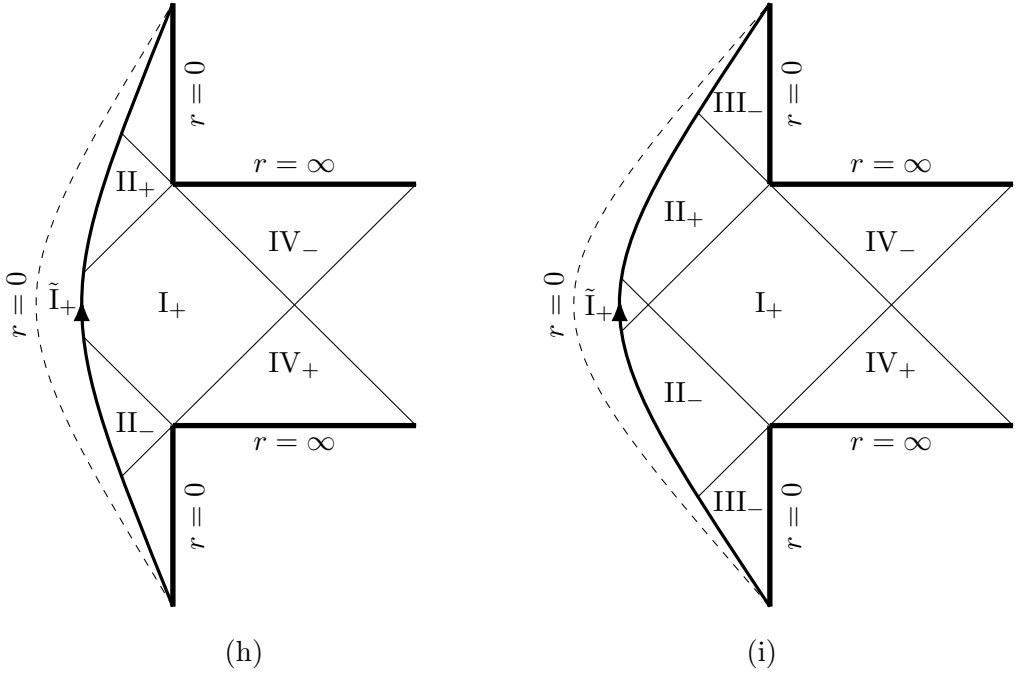
\begin{figure}
\centering
\begin{minipage}{.42\textwidth}
\begin{tikzpicture}[scale=0.8]
\draw[line width=2pt] (4,-1)--(4,2)node[midway,below, sloped] (TextNode) {$r=0$};
\draw[line width=2pt] (4,2)--(8,2)node[midway,below, sloped] (TextNode) {$r=\infty$};
\draw[line width=2pt] (4,6)--(8,6)node[midway,above, sloped] (TextNode) {$r=\infty$};
\draw[line width=2pt] (4,6)--(4,9)node[midway,below, sloped] (TextNode) {$r=0$};
\draw (6,4)--(8,6);
\draw (6,4)--(8,2);
\draw (6,4)--(4,6);
\draw (6,4)--(4,2);
\draw[very thick] (4,-1)..controls (2,4)..(4,9);
\draw[dashed] (4,-1)..controls (1,4)..(4,9)node[midway,above, sloped] (TextNode) {$r=0$};
\draw[->,>=Latex, very thick](2.5,4)--(2.5,4.1);
\draw (4,2)--(3.15,1.15);
\draw (4,2)--(2.55,3.45);
\draw (4,6)--(3.15,6.85);
\draw (4,6)--(2.55,4.55);
\node at (4,4){$\rI_+$};
\node at (6,3){$\rIV_+$};
\node at (6,5){$\rIV_-$};
\node at (3.4,6){$\rII_+$};
\node at (3.4,2){$\rII_-$};
\node at (2.2,4.1){$\tilde \rI_+$};
\end{tikzpicture}
\captionof*{figure}{(h)}
\end{minipage}
\begin{minipage}{.42\textwidth}
\begin{tikzpicture}[scale=0.8]
\draw[line width=2pt] (4,-1)--(4,2)node[midway,below, sloped] (TextNode) {$r=0$};
\draw[line width=2pt] (4,2)--(8,2)node[midway,below, sloped] (TextNode) {$r=\infty$};
\draw[line width=2pt] (4,6)--(8,6)node[midway,above, sloped] (TextNode) {$r=\infty$};
\draw[line width=2pt] (4,6)--(4,9)node[midway,below, sloped] (TextNode) {$r=0$};
\draw (6,4)--(8,6);
\draw (6,4)--(8,2);
\draw (6,4)--(4,6);
\draw (6,4)--(4,2);
\draw[very thick] (4,-1)..controls (0.7,4)..(4,9);
\draw[dashed] (4,-1)..controls (-0.3,4)..(4,9)node[midway,above, sloped] (TextNode) {$r=0$};
\draw[->,>=Latex, very thick](1.53,4)--(1.53,4.1);
\draw (4,2)--(2.8,0.8);
\draw (4,2)--(2,4);
\draw (4,6)--(2.8,7.2);
\draw (4,6)--(2,4);
\draw (2,4)--(1.55,3.55);
\draw (2,4)--(1.55,4.45);
\node at (4,4){$\rI_+$};
\node at (6,3){$\rIV_+$};
\node at (6,5){$\rIV_-$};
\node at (2.6,5.5){$\rII_+$};
\node at (2.6,2.5){$\rII_-$};
\node at (3.5,7.3){$\rIII_-$};
\node at (3.5,0.7){$\rIII_-$};
\node at (1.2,4){$\tilde \rI_+$};
\end{tikzpicture}
\captionof*{figure}{(i)}
\end{minipage}
\caption{\label{fig:h_i}For trajectories h) and i), the turning points also lie in region $\rI$, but the trajectories start at $R = 0$. The expanding shells reach the inner turning point in region $\rI_+$ (path h)) or $\rI_-$ (path i)) depending on the sign of $\eps_+$ at this radius. The inside is described by region $\tilde \rI_+$ of de Sitter space.}
\end{figure}

Trajectories h) and i) are sketched in Fig.~\ref{fig:h_i}. In both cases, the trajectory starts at $R = 0$ with $\eps_+ = -1$ (region $\rIII_-$), passes through region $\rII$ into $\rI$, where it has a turning point, and then continues again to $\rIII_-$. In the case h), the turning point is greater than $R_+$, i.e., it lies in $\rI_+$, while in the case i), $\eps_+ = -1$ throughout, so the turning point lies in region $\rI_-$. In case h), the shell is seen to leave the ``white hole'' region, but collapses again into the black hole, ending up in the singularity. In case i), the shell does not leave the white hole region, so the exterior can be interpreted as an ``eternal'' black hole. Analogous trajectories also exist for vanishing cosmological constant, \cf case 3a) in \cite{Boulware1973}, and for vanishing charge $Q = 0$ \cite{YamanakaEtAl}.

Finally, for trajectory j), shown in Fig.~\ref{fig:j}, $\eps_+$ changes sign in $\rIV$, so a collapsing shell passes through regions $\rI_-$ and $\rIII_-$ before crashing into the singularity. Such a trajectory does not exist for vanishing cosmological constant \cite{Boulware1973}, but is also present for vanishing charge $Q = 0$ \cite{YamanakaEtAl}.

\begin{figure}
\centering
\begin{minipage}{.42\textwidth}
\begin{tikzpicture}[scale=0.8]
\draw[line width=2pt] (2.15,0)--(4,0)node[midway,below, sloped] (TextNode) {$r=\infty$};
\draw[line width=2pt] (8,7.8)--(8,4)node[midway,above, sloped] (TextNode) {$r=0$};
\draw (4,0)--(6,2);
\draw (6,2)--(8,4);
\draw (2.15,0)--(2.95,0.8);
\draw[very thick] (2.5,0)..controls (4.5,3.5)..(8,7.8);
\draw[dashed] (2.15,0)..controls (4.4,4)..(8,7.8)node[midway,above, sloped] (TextNode) {$r=0$};
\draw[->,>=Latex, very thick](5.3,4.4)--(5.38,4.5);
\draw (4,0)--(3.05,0.95);
\draw (8,4)--(6.3,5.7);
\draw (6,2)--(4.55,3.45);
\node at (4.5,1.5){$\rI_-$};
\node at (6.2,3.7){$\rII_+$};
\node at (7.4,5.6){$\rIII_-$};
\end{tikzpicture}
\end{minipage}
\hspace{.2cm}
\begin{minipage}{.42\textwidth}
\begin{tikzpicture}[scale=0.8]
\draw[line width=2pt] (2.6,8)--(4,8)node[midway,above, sloped] (TextNode) {$r=\infty$};
\draw[line width=2pt] (8,4)--(8,0.2)node[midway,above, sloped] (TextNode) {$r=0$};
\draw[very thick] (8,0.2)..controls (4.2,5)..(3,8);
\draw[dashed] (8,0.2)..controls (4,4.6)..(2.6,8)node[midway,below, sloped] (TextNode) {$r=0$};
\draw[->,>=Latex, very thick](4.9,4.2)--(4.82,4.3);
\draw (8,4)--(6,6);
\draw (6,6)--(4,8);
\draw (8,4)--(6.32,2.32);
\draw (6,6)--(4.62,4.62);
\draw (4,8)--(3.3,7.3);
\draw (3.25,7.35)--(2.6,8);
\node at (7.5,2){$\rIII-$};
\node at (6.3,4.2){$\rII_-$};
\node at (4.5,6.2){$\rI_-$};
\end{tikzpicture}
\end{minipage}
\captionof*{figure}{(j)}
\caption{\label{fig:j}Path j) may describe a collapse (left) or an explosion (right) depending on the sign of $\dot{R}$. As $\eps_+$ changes sign in regions of type $\rIV$, the shell trajectory must evolve through region $\rI_-$.}
\end{figure}

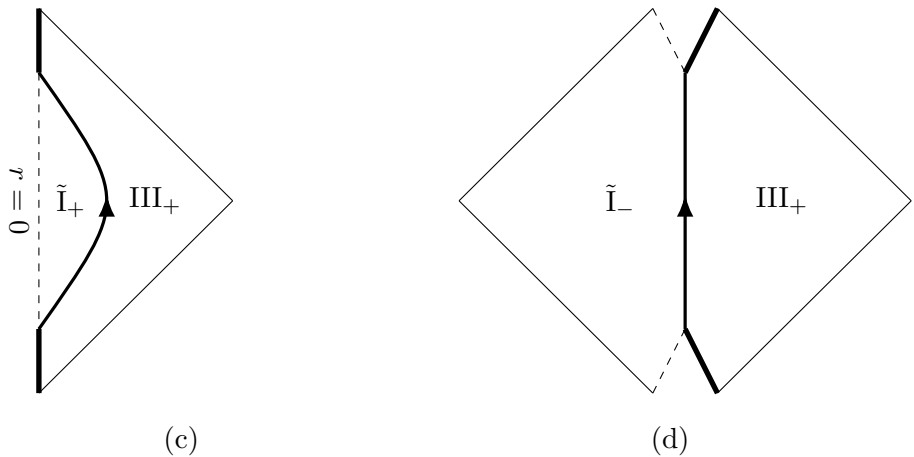
\begin{figure}
\centering
\begin{minipage}{.3\textwidth}
\begin{tikzpicture}[scale=0.85]
\draw [dashed](6,5)--(6,1)node[midway,below, sloped] (TextNode) {$r=0$};
\draw[very thick] (6,1)..controls (7.4,3)..(6,5);
\draw[->,>=Latex, very thick](7.05,3)--(7.05,3.1);
\draw[line width=2pt] (6,6)--(6,5);
\draw[line width=2pt] (6,1)--(6,0);
\draw (9,3)--(6,0);
\draw (9,3)--(6,6);
\node at (7.8,3){$\rIII_+$};
\node at (6.5,3){$\tilde \rI_+$};
\end{tikzpicture}
\captionof*{figure}{(c)}
\end{minipage}
\hspace{1cm}
\begin{minipage}{.35\textwidth}
\begin{tikzpicture}[scale=0.85]
\draw[very thick] (5.5,1) -- (5.5,5);
\draw[->,>=Latex, very thick](5.5,3)--(5.5,3.1);
\draw[line width=2pt] (6,6)--(5.5,5);
\draw[line width=2pt] (6,0)--(5.5,1);
\draw[dashed] (5,6)--(5.5,5);
\draw[dashed] (5,0)--(5.5,1);
\draw (9,3)--(6,0);
\draw (9,3)--(6,6);
\draw (2,3)--(5,0);
\draw (2,3)--(5,6);
\node at (7,3){$\rIII_+$};
\node at (4.5,3){$\tilde \rI_-$};
\end{tikzpicture}
\captionof*{figure}{(d)}
\end{minipage}
\caption{\label{fig:c_d_neg_M}The trajectories c) and d) for the case of $M < 0$.}
\end{figure}

We may now consider the case $M < 0$ of negative rest mass. As both $R_\pm$ and the potential $V(R)$ depend on $M$ only through $M^2$, the position of the turning points and of the radii $R_\pm$ does not change under the replacement $M \to -M$. In particular, Figs.~\ref{fig:Q<m} and \ref{fig:Q>m} are also valid under this replacement. The signs $\eps_\pm$, however, are inverted under $M \to -M$. As an example, Fig.~\ref{fig:c_d_neg_M} shows the form of the trajectories c) and d) in this case.

Up to now, we discussed the generic black hole case. Trajectories for the extremal cases are discussed in \cite{Giese}. We finish by discussing the possibility of creating a naked singularity by collapse of a charged spherical dust shell. Hence, we are looking for a trajectory ending in the singularity $r = 0$ of region $\rI_+$ of the generic naked singularity case, \cf Fig.~\ref{fig:NakedSingularity}. We thus require $\eps_+ = +1$ for $r < r_c$. Furthermore, we want the inside of the shell to be bounded, i.e., we require also $\eps_- = +1$, so that the interior of the shell is described by region $\tilde \rI_+$, \cf Fig.~\ref{fig:dS}.

\begin{figure}
\centering
\includegraphics{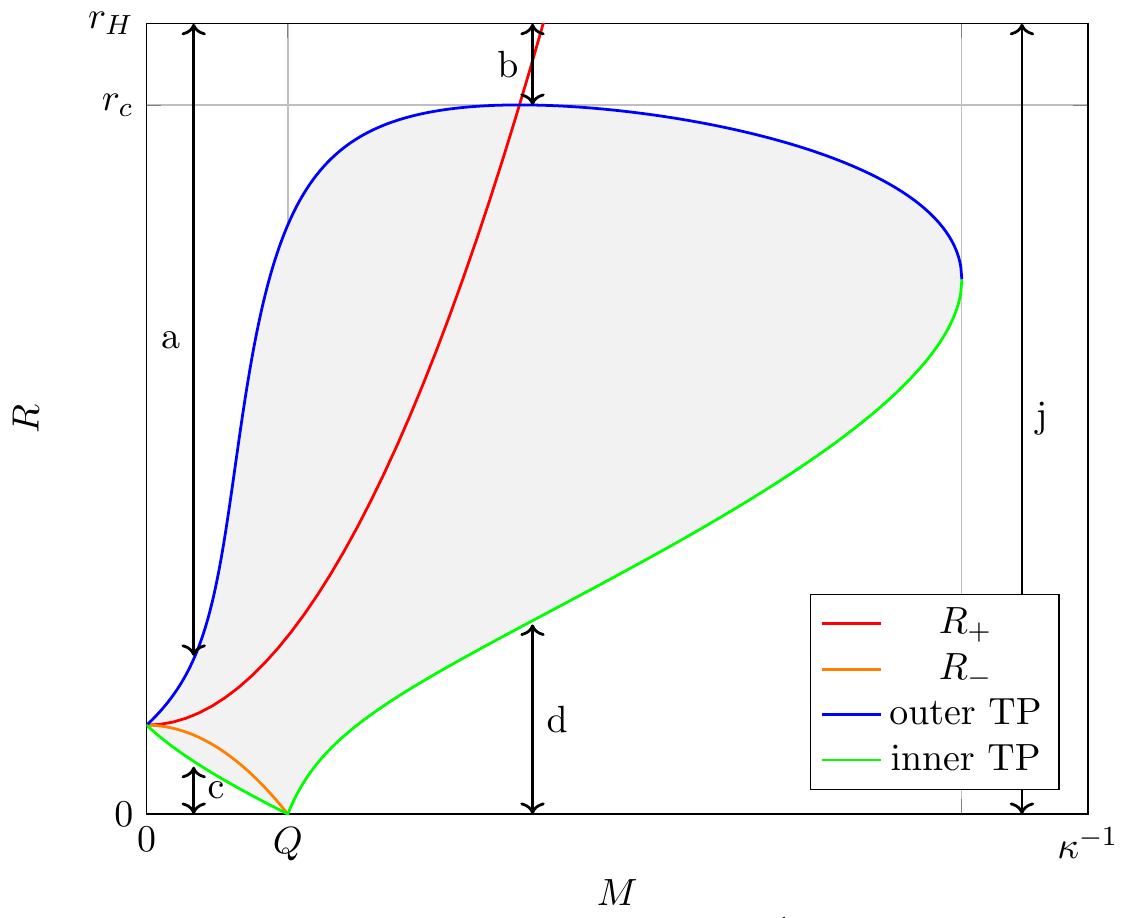}
\caption{\label{fig:NakedSingularityPhaseDiagram}Same as Fig.~\ref{fig:Q<m} but for the naked singularity case with $Q > m$ ($m = 0.1 \kappa^{-1}$, $Q = 0.15 \kappa^{-1}$).}
\end{figure}

We now need to know the position of the turning points. The analysis performed above for the generic black hole case is still valid, i.e., while in region $\rI$, both $R_+$ and $R_-$ are in the forbidden region between two turning points, and $R_-$ becomes negative at $Q^2 = M^2$. This is also apparent in Fig.~\ref{fig:NakedSingularityPhaseDiagram}, which shows $R_\pm$ and the turning points as a function of $M$ for $m < Q$. In contrast to the generic black hole case, there is now only a single forbidden region, contained in $\rI$, whose shape resembles that of the forbidden region of $\rIII$ in the near extremal case, Fig.~\ref{fig:Q>m}. Hence, there is a reduced number of types of trajectories. In the naked singularity case with $m > Q$, the trajectory of type d) is absent, as the curve describing the outer turning point terminates at $M = Q$ (as for the forbidden region contained in $\rIII$ in the sub-extremal case shown in Fig.~\ref{fig:Q<m}).


As is obvious from Fig.~\ref{fig:NakedSingularityPhaseDiagram}, in the naked singularity case the only trajectories which reach the singularity are c), d), and j). The last two have signs $(\eps_+, \eps_-) = \sign(M) (-1, +1)$ when reaching the singularity (recall that under the change $M \to -M$, Fig.~\ref{fig:NakedSingularityPhaseDiagram} remains valid, but the signs $(\eps_+, \eps_-)$ are flipped). They are thus not of the desired form to describe collapse to a naked singularity. On the other hand, the trajectory c) has signs $(\eps_+, \eps_-) = \sign(M) (-1, -1)$ when reaching the singularity, so it has the desired form for negative rest mass $M$. This trajectory is like the one depicted on the \lhs of Fig.~\ref{fig:c_d_neg_M}, with $\rIII_+$ relabelled as $\rI_+$. This trajectory also has a counterpart for vanishing cosmological constant, case 4) in \cite{Boulware1973}. As for a vanishing cosmological constant \cite{Boulware1973}, we can thus conclude that collapse to a naked singularity is only possible for negative rest mass of the shell.


For completeness, we mention that the first sub-case 3b) of \cite{Boulware1973} corresponds to our trajectory d) in the naked singularity case with $M > 0$ (the second sub-case of 3b) of \cite{Boulware1973} is empty for $m > 0$).

\section{Conclusion}

We have found that all types of trajectories that are present for vanishing cosmological constant \cite{Boulware1973} are also present for a positive cosmological constant. In particular, there are the trajectories a), b) and e) describing the formation of a black hole. We have proven that for all parameters of the generic black hole case the ``phase diagram'' of the turning points is as represented in Figs.~\ref{fig:Q<m} and \ref{fig:Q>m} for the cases $m>Q$ and $m<Q$. Hence, for any set of parameters $(m, Q, \Lambda)$ describing a RNdS black hole, there is a charged shell configuration which collapses to such a black hole. As there are spacetimes in this class for which the strong cosmic censorship conjecture is violated \cite{CardosoEtAl}, this means that gravitational collapse spacetimes exist for which the strong cosmic censorship conjecture is violated. (This is not a contradiction to the claim made in \cite{Hod2019} that such spacetimes can not form by gravitational collapse, as our analysis is purely classical, while for the argument put forward in \cite{Hod2019} quantum (gravity) effects are essential.)


We also saw that for a non-vanishing cosmological constant also new types of shell trajectories are possible, in particular trajectories which bounce before the formation of a horizon (analogous trajectories are also present in the vanishing charge case $Q = 0$ \cite{YamanakaEtAl}). Finally, we showed that collapse to a naked singularity is only possible for negative rest mass $M$ of the shell, i.e., in a situation typically excluded by energy conditions.

\end{document}